\documentclass[a4paper,fleqn,usenatbib]{mnras}

\usepackage[T1]{fontenc}
\usepackage{ae,aecompl}

\usepackage{graphicx}	
\usepackage{amsmath}	
\usepackage{amssymb}	
\usepackage{threeparttable} 
\usepackage{xcolor}

\title[PSR J1012+5307]{PSR J1012+5307: a millisecond pulsar with an extremely low-mass white dwarf companion}
 
\author[Mata S\'{a}nchez et al.] {D. Mata S\'{a}nchez$^{1}$\thanks{E-mail: matasanchez.astronomy@gmail.com}, A. G. Istrate$^{2}$, M. H. van Kerkwijk$^{3}$, R. P. Breton$^{1}$, 
\newauthor D. L. Kaplan$^{4}$ 
\\
\\
$^{1}$Jodrell Bank Centre for Astrophysics, Department of Physics and Astronomy, The University of Manchester, M13 9PL, UK
\\
$^{2}$Department of Astrophysics/IMAPP, Radboud University, PO Box 9010, NL-6500 GL Nijmegen, The Netherlands
\\
$^{3}$Department of Astronomy and Astrophysics, University of Toronto, 50 St. George Street, Toronto, ON M5S 3H4, Canada
\\
$^{4}$Center for Gravitation, Cosmology and Astrophysics, Department of Physics, University of Wisconsin-Milwaukee, PO Box 413,
\\Milwaukee, WI 53201, USA
}
 
\date{Accepted 2020 April 03. Received 2020 April 03; in original form 2020 March 10 }

\pubyear{2020}

\begin{document}
\label{firstpage}
\pagerange{\pageref{firstpage}--\pageref{lastpage}}
\maketitle

\begin{abstract}
 Binaries harbouring millisecond pulsars enable a unique path to determine neutron star masses: radio pulsations reveal the motion of the neutron star, while that of the companion can be characterised through studies in the optical range. PSR J1012+5307 is a millisecond pulsar in a $14.5$-h orbit with a helium-core white dwarf companion. In this work we present the analysis of an optical spectroscopic campaign, where the companion star absorption features reveal one of the lightest known white dwarfs. We determine a white dwarf radial velocity semi-amplitude of $K_2 = 218.9 \pm 2.2\, \rm km\,s^{-1}$, which combined with that of the pulsar derived from the precise radio timing, yields a mass ratio of $q=10.44\pm 0.11$. We also attempt to infer the white dwarf mass from observational constraints using new binary evolution models for extremely low-mass white dwarfs, but find that they cannot reproduce all observed parameters simultaneously. In particular, we cannot reconcile the radius predicted from binary evolution with the measurement from the photometric analysis ($R_{\rm WD}=0.047_{-0.002}^{+0.003}\, R_{\odot}$). Our limited understanding of extremely low-mass white dwarf evolution, which results from binary interaction, therefore comes as the main factor limiting the precision with which we can measure the mass of the white dwarf in this system. Our conservative white dwarf mass estimate of $M_{\rm WD} = 0.165 \pm 0.015\, M_{\rm \odot}$, along with the mass ratio enables us to infer a pulsar mass of $M_{\rm NS} = 1.72 \pm 0.16\, M_{\rm \odot}$. This value is clearly above the canonical $\sim 1.4\, M_{\rm \odot}$, therefore adding PSR J1012+5307 to the growing list of massive millisecond pulsars.
\end{abstract}




\section{Introduction}
\label{introduction}

Millisecond pulsars (hereafter MSPs) are an extreme class of rapidly spinning radio pulsars. Since their discovery almost $\sim 40$ years ago \citep{Backer1982} more than $\sim 400$ MSPs (accounting for $\sim 15 \%$ of the pulsars population) have been found so far (e.g., \citealt{Manchester2005}). Their elevated spin frequencies are too high to be acquired at birth, and instead require the MSP to have spun up due to mass transfer from a companion star \citep{Radhakrishnan1982}. 
As a result of their evolutionary history, MSPs are expected to harbour neutron stars (NSs) more massive than their counterparts of the slowly rotating pulsar population. This claim has been supported by the discovery of massive NSs among the MSP kind (see e.g., \citealt{Ozel2016}). Therefore, the study of the MSP population is crucial to empirically determine the maximum allowed mass for NSs, one of the key constraints in obtaining the correct equation of state (as each of them predict different theoretical mass limits, see \citealt{Lattimer2012} for a review). 

Mass measurements of NSs are traditionally only feasible in binary systems, where the timing analysis of radio pulsations allows us to trace the NS movement. Radio timing is enough to unambiguously determine the NS mass only for the handful of systems where two or more post-Keplerian parameters may be measured such as Shapiro delay (e.g., \citealt{Demorest2010}), precession of periastron (e.g., \citealt{Freire2008}), or others (see \citealt{Lorimer2012}). For the remaining NS-harbouring binaries, multiwavelength campaigns are required to  obtain their dynamical solution (i.e. the NS mass), as the companion star movement is usually characterised via optical/near-infrared spectroscopy.

PSR J1012+5307 (J1012 hereafter) was originally discovered by \citet{Nicastro1995}, and identified as a MSP pulsar in a binary system. Initial studies based on optical spectroscopy (\citealt{vanKerkwijk1996}, \citealt{Callanan1998}) revealed a extremely light white dwarf (WD) companion star ($\sim 0.16\, M_{\rm \odot}$), establishing J1012 as a member of the extremely low-mass WDs class (hereafter ELM WDs, e.g., \citealt{Marsh1995,Brown2013}). ELM WDs are though to be born in binary systems with a sufficiently compact orbit, where the sub-giant WD progenitor envelope is stripped before they reach the red giant phase, leaving behind a light He-core WD (e.g., \citealt{Tauris1996a}, \citealt{Tauris1996b}, \citealt{Istrate2014}). 

In this work, we have unveiled the radial velocity of the WD in J1012 with an unprecedented precision using optical spectroscopy of the highest spectral resolution performed to date. Combined with other derived parameters for the WD, both within this work and in the literature, it allows us to derive the dynamical solution of the system, as well as to discuss its evolutionary history.

\section{Observations}
\label{observations}

We observed J1012 using both the Keck I and Keck II 10-m telescopes\footnote{The data presented herein were obtained at the W. M. Keck Observatory on Mauna
Kea, Hawaii, which is operated as a scientific partnership among the California Institute of Technology, the University of California and the National Aeronautics and Space Administration. The Observatory was made possible by the generous financial support of the W. M. Keck Foundation. } (Hawaii, USA), respectively equipped with the Low-Resolution Imaging Spectrometer (LRIS, \citealt{Oke1995}; producing 23 spectra) and the Echellette Spectrograph and Imager (ESI, \citealt{Sheinis2002}; yielding 16 spectra). The observation log describing our complete spectral dataset (covering from 1995 to 2005) is detailed in Table \ref{tab:Obslog}. 

 Out of the 23 LRIS spectra (obtained during the period 1995-1997), 22 cover the spectral range of 3700 -- 6000 \AA , using a low-resolution grism and two different slit widths (0.7'' and 1'', depending on the daily observing conditions). This setup results in typical spectral resolutions of $ \sim 190\, {\rm km\, s^{-1}}$ and $ \sim 260\, {\rm km\, s^{-1}}$, respectively (see Tab. \ref{tab:Obslog}). A single spectrum of higher spectral resolution ($ \sim 100\, {\rm km\, s^{-1}}$) but covering a narrower wavelength range centred at the H$\rm \alpha$ line (5800--7200 \AA) was also obtained as a result of the available instrumental setup for one particular night. All these spectra are reduced using semiautomatic routines developed by our team based on \textsc{iraf}\footnote{IRAF is distributed by the National Optical Astronomy Observatories, operated by the Association of Universities for Research in Astronomy, Inc., under contract with the National Science Foundation.}, \textsc{molly}\footnote{\textsc{molly} software developed by T. R. Marsh.} and \textsc{python}. Each spectrum is de-biased and flat-field corrected. Arcs were obtained before and/or after each observing block. To perform a precise wavelength calibration of each target spectrum, we select the arc obtained closer in time, extracting an individual arc spectrum from the two-dimensional image at the corresponding position defined by the target aperture. The wavelength calibration is refined comparing sky emission lines (in particular \ion{O}{i}--$\lambda$5577) with their corresponding rest wavelengths, from which we derive sub-pixel velocity drifts that are subsequently corrected. Finally, we correct the spectra for the relative velocity between the Earth and the solar system barycentre at each observational epoch.

The ESI echelle spectrograph provided us with 16 spectra (obtained during 3 consecutive nights in 2005) of higher spectral resolution and wider wavelength range (useful range 4000--10000 \AA, $ \sim 45-60\, \rm{km\, s^{-1}}$ depending on the slit width). They are reduced in a similar way to the LRIS data, producing for each exposure 10 spectra corresponding to the different spectral orders of the echellette spectrograph. Each exposure is then blaze-corrected using the spectral flats continuum, and they are finally merged (after performing a weighted average in the overlapping regions) into a single spectrum covering the full wavelength range. 

We flux-calibrate all the spectra using the standard star Feige 34 as reference, which was observed each night with the same setup as J1012. We also account for atmospheric extinction combining the airmass value during the observations with the tables provided for Mauna Kea by \citet{MaunaKea1988}. Nevertheless, we note that the final flux calibration must be taken with caution, due to effects such as slit losses or the presence of clouds during some particular nights. We also produce a normalised version of the spectra dividing them by a low-order polynomial fit of the continuum. We note that the normalisation of the ESI spectra produced better results when performed on each spectral order independently before merging them into a single spectrum, so we decided to employ these for further analysis.

\begin{table*}
\caption{Observation log: it includes the barycentric Julian date (BJD) at mid exposure time for each observation, the telescope/instrument employed, the instrumental setup (including the selected grism for LRIS, and the slit width in all cases), the spectral resolution (R, measured through a Gaussian fit to the sky lines), the exposure time ($ T_{\rm{EXP}}$) and the radial velocity of the WD measured in each spectrum using the cross-correlation technique described in Sec. \ref{RV}. } 

\bigskip

\centering
\begin{threeparttable}
\begin{tabular}{l l l l l l r}
\hline
BJD (d) & Tel./Inst.$^{\rm a}$          & Setup - Slit & R $\,(\rm{km\,s^{-1}})$ & Airmass &   $ T_{\rm{EXP}} \,(\rm{s})$  & $v\, (\rm{km\, s^{-1})}$ \\
\hline
2450044.04075856  &	Keck I/LRIS	& 600/5000 -	0.7'' &		$185$		& 1.73 &		1800 & $-87 \pm 15$	  \\ 
2450044.15022052	  &	Keck I/LRIS	&  600/5000 -	0.7'' &	$185$		&	1.23 &	1080 & $154 \pm 17$	   \\ 
2450046.1319027 	 &	Keck I/LRIS	& 600/5000	 -	1.0''	&	$260$		&	 1.25 &	1200 & $74 \pm 27$	   \\
       2450046.14720385	 &	Keck I/LRIS	& 600/5000	 -	1.0''	&		$260$	&	1.23 &	1200 & $25 \pm 16$	   \\ 
       2450047.08355071 &	Keck I/LRIS	& 600/5000	  -	1.0'' &		$260$		&	1.37 &	358$^{\rm b}$	 & $29 \pm 51$   \\ 
       2450047.09721602 &	Keck I/LRIS	& 600/5000	  -	0.7'' &		$180$		&	1.35 &	1800 & $24 \pm 23$	   \\ 
       2450103.03720118	 &	Keck I/LRIS	& 600/5000	  -	0.7'' &		$180$		&	1.20 &	821$^{\rm b}$ & $-80 \pm 24$	   \\ 
       2450103.10721253	 &	Keck I/LRIS	& 600/5000	  -	0.7''  &	$180$		&	1.27 &	1800 & $-161 \pm 10$	   \\ 
       2450103.131563	 &	Keck I/LRIS	& 600/5000	  -	0.7''  &	$180$			& 1.34 &		1800 & $-218 \pm 10$	   \\  
       2450103.1515573	 &	Keck I/LRIS	& 600/5000	  -	0.7''  &	$180$	 	&	1.42 &	1500 & $-204 \pm 11$	   \\  
       2450103.16805153	 &	Keck I/LRIS	& 600/5000	  -	0.7''  &	$180$			& 1.52	&	1200 & $-215 \pm 24$	   \\  
       2450103.9895485	 &	Keck I/LRIS	& 600/5000	  -	0.7'' &		$180$		&	1.24 &	1800 & $132 \pm 8$	   \\ 
       2450104.01439896	 &	Keck I/LRIS	& 600/5000	  -	0.7''  &	$180$			&  1.21	&	1800	 & $160 \pm 8$   \\ 
       2450428.99356788	 &	Keck I/LRIS	& 600/5000	  -	0.7'' &		$200$	& 1.69	&		1800 & $-155 \pm 11$	   \\  
       2450429.01906951	 &	Keck I/LRIS	& 600/5000	  -	0.7''  &	$200$			& 1.51	&	1800 & $-188 \pm 8$	   \\   
       2450429.07079282	 &	Keck I/LRIS	& 600/5000	  -	0.7''  &	$200$			&	1.30 &	1800 & $-230 \pm 10$	   \\   
       2450429.09483435	 &	Keck I/LRIS	& 600/5000	  -	0.7''  &	$200$			&	1.24 &	1800 & $-224 \pm 9$	   \\  
       2450429.13966721	 &	Keck I/LRIS	& 600/5000	  -	0.7''   &	$200$			&	1.20 &	1800 & $-189 \pm 11$	   \\   
       2450429.16049653	 &	Keck I/LRIS	& 600/5000	  -	0.7''   &	$200$			&	1.20 &	1200 & $-167 \pm 10$	   \\   
       2450430.02959768	 &	Keck I/LRIS	& 600/5000	  -	1.0'' &		$270$	& 1.44	&		1749$^{\rm b}$	 & $169 \pm 30$   \\   
       2450430.07537325	 &	Keck I/LRIS	&  600/5000	  -	1.0''  &	$270$			& 1.26	&	977$^{\rm b}$	 & $92 \pm 19$   \\ 
       2450457.16393162	 &	Keck I/LRIS	& 600/5000	  -	0.7'' &		$210$	& 1.36	&		1200 & $186 \pm 13$	   \\ 
       2450793.14717118	 &	Keck I/LRIS	& 1200/7500	  -	1.0'' &		$110$	& 1.20	&		1500 & $-213 \pm 8$	   \\  
       2453434.80431978	 &	Keck II/ESI	&  	0.75'' &		$45$	& 1.45	&		1200	  & $-103 \pm 6$  \\  
       2453434.83417997	 &	Keck II/ESI	&  	0.75'' &		$45$		 		&	1.36	& 1800 & $-160 \pm 5$	   \\  
       2453434.85649811	 &	Keck II/ESI	&  	0.75'' &		$45$		 		&	1.29 &	1800 & $-194 \pm 4$	   \\  
       2453434.88057517	 &	Keck II/ESI	&  		0.75'' &		$45$		 		&	1.23 &	1800 & $-221 \pm 5$	   \\   
       2453434.90220433	 &	Keck II/ESI	&  		0.75'' &		$45$		 		&	1.21 &	1800 & $-233 \pm 5$	   \\  
       2453434.99584265	 &	Keck II/ESI	&  		0.75'' &		$45$		 		&	1.27 &	1800 & $-189 \pm 5$	   \\   
       2453435.75020639	 &	Keck II/ESI	&  	0.75'' &		$45$	& 1.80$^{\rm c}$	&		1800 & $160 \pm 22$	   \\   
       2453435.78031156	 &	Keck II/ESI	&  		0.75'' &		$45$				&	1.56$^{\rm c}$ &	1800 & $175 \pm 17$	   \\   
       2453436.04166144	 &	Keck II/ESI	&  		0.75'' &		$45$				&	1.49$^{\rm c}$ &	1800 & $-179 \pm 7$	   \\    
       2453436.74927761  &	Keck II/ESI	&  	1.0'' &		$60$	& 1.78	&		1800 & $-223 \pm 10$	   \\    
       2453436.77695808 &	Keck II/ESI	&  	1.0'' &		$60$				&	1.56 & 	1800 & $-185 \pm 12$	   \\  
	   2453436.80358305 &	Keck II/ESI	&  	1.0'' &		$60$				&	1.43 &	2400 & $-174 \pm 6$	   \\ 
       2453436.83366445 &	Keck II/ESI	&  	1.0'' &		$60$				&	1.31 &	2400 & $-111 \pm 5$	   \\  
       2453436.86479107 &	Keck II/ESI	&  		1.0'' &		$60$				&	1.24 &	2400 & $-54 \pm 4$	   \\  
       2453436.90312919 &	Keck II/ESI	&  		1.0'' &		$60$				&	1.20 &	2400 & $40 \pm 5$	   \\  
       2453436.97147577 &	Keck II/ESI	&  		1.0'' &		$60$				&	1.24 &	2400 & $164 \pm 4$	   \\  
\hline
\end{tabular}
\begin{tablenotes}
\item[a]{The Keck Observatory telescopes (Keck I and II), equipped with either the Low-Resolution Imaging Spectrometer (LRIS) or the Echellette Spectrograph and Imager (ESI).}
\item[b]{The observation was interrupted due to bad weather/technical issues, resulting in the reported exposure time.}
\item[c]{The presence of clouds during the observing run reduced significantly the signal-to-noise of the retrieved spectra.}
\end{tablenotes}
\label{tab:Obslog}
\end{threeparttable}
\end{table*}

\section{Results}
\label{results}

The spectra are dominated by the broad absorption lines of the hydrogen Balmer series produced at the WD companion atmosphere. No other features (apart from telluric lines) are found, in neither the low-resolution nor the high-resolution spectra.

\subsection{Radial velocity of the WD companion}
\label{RV}

To trace the WD movement along the orbit, we measure the radial velocity shifts of the Balmer absorption lines in the normalised spectra at different orbital phases. For this purpose, we apply cross-correlation techniques to compare J1012 spectra with synthetic WD templates (obtained from \citealt{Koester2008}). We compare each observed spectrum with a template of effective temperature $T_{\rm eff}=8500\, {\rm K}$ and surface gravity $\log g = 6.5 $ (as reported by \citealt{vanKerkwijk1996,Callanan1998}). We also mask out all the features unrelated with the WD companion, such as telluric lines and reduction artifacts. We note that the masks are different for the two instruments, as they cover distinct wavelength regimes: while the lower resolution LRIS spectra allow us to inspect the Balmer series from $\rm H\beta$ to $\rm H12$, ESI spectra allow us to access $\rm H\alpha$ to $\rm H\delta$ transitions. We then use the \textit{crosscorrRV} PyAstronomy\footnote{https://github.com/sczesla/PyAstronomy} routine to shift the template spectrum along a range of velocities ($-1000$ to $1000\, \rm km \, s^{-1}$ in steps of $10\, \rm{km \, s^{-1}}$), performing the cross-correlation with the object spectrum for each velocity shift. This generates, for each observed spectrum, a cross-correlation function that peaks at the actual WD radial velocity. We determine such peak via a parabolic fit to the maximum, which produce purely statistical uncertainties, overestimating our precision in determining the radial velocities. In order to obtain more realistic uncertainties on the derived radial velocities, we perform a Monte Carlo analysis. For each object spectrum we simulate 1000 spectra, using the observed spectrum as a seed and assuming that the measured flux and wavelength values follow Gaussian distributions with standard deviations determined by the flux uncertainty and the root-mean-square of the wavelength calibration, respectively. The application of the cross-correlation technique previously described to the simulated spectra allow us to obtain a distribution of radial velocities for each observed spectrum. We finally determine the radial velocity associated with each spectrum from the mean and standard deviation of its corresponding distribution.

\begin{figure*}
\includegraphics[width=\columnwidth,trim={1.5cm 0 2cm 0},clip]{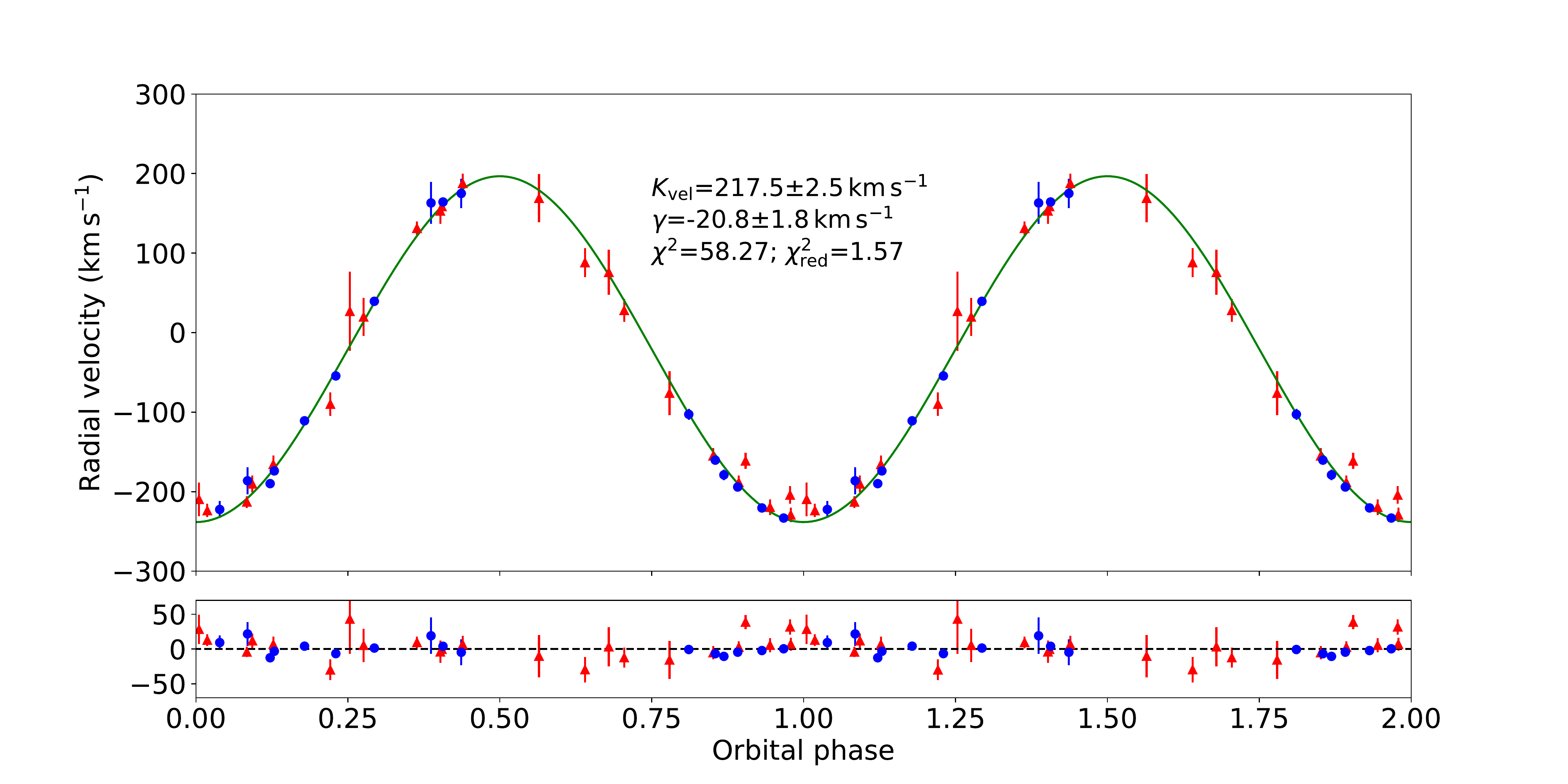}
\includegraphics[width=\columnwidth,trim={1.5cm 0 2cm 0},clip]{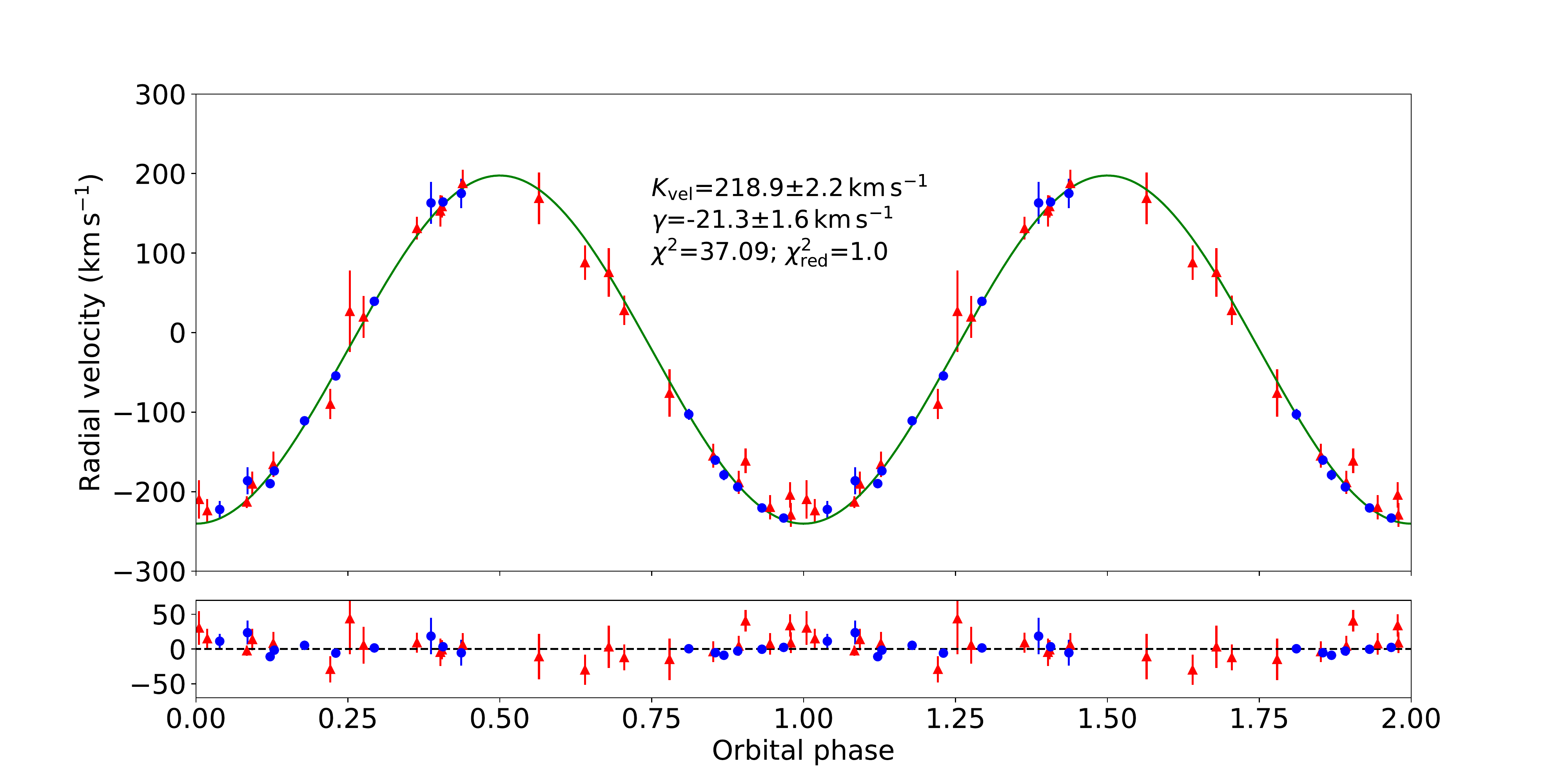}
\caption{Left panel: Radial velocity curve obtained through cross-correlation of the individual spectra with a synthetic WD template ($T_{\rm eff}=8500\, {\rm K}$, $\log g = 6.50$). Blue circles correspond to the high resolution ESI data. Red triangles depict the lower resolution dataset obtained with LRIS. A sinusoidal fit corresponding to the circular orbit motion is depicted as a green, continuous line. Two orbital phases are shown for clarity. Note that the null orbital phase corresponds to $T_{\rm asc}$. Right panel: same as the left panel, but including an extra systemic contribution to the uncertainties of LRIS radial velocities in order to make $\chi_{\rm red}^{2}=1$.}
\label{figK2rv}
\end{figure*}

As each observation was scheduled to cover a different orbital phase, we can construct the radial velocity curve for the WD (see Fig. \ref{figK2rv}). We fit the measured radial velocities with an orbit with all parameters except the systemic velocity and velocity amplitude fixed to the most accurate elements available: the orbital eccentricity ($e=0$ to our precision), the orbital period ($P_{\rm{orb}}= 0.60467271355\pm 0.00000000003\, \rm{d} $) and time of the ascending node ($T_{\rm{asc}}=2450700.58162891\pm 0.00000004\, \rm{d}$, \citealt{Lazaridis2009}), which corresponds to the null orbital phase in our reference frame. The resulting fit of our complete dataset (Fig. \ref{figK2rv}, left panel) yields a WD radial velocity of $K_2= 217.5 \pm 2.5 \, \rm{km \, s^{-1}}$ and a systemic velocity (not including a small correction for gravitational redshift of the white dwarf; see Sec. \ref{dist3Dvel}) of $\gamma = -20.8\pm 1.8 \, \rm{km \, s^{-1}}$ ($\chi^2_{\rm{red}}=1.57$ for 37 d.o.f.). 

To asses the reliability of the derived parameters, we first note that the lower resolution LRIS spectra were not observed with a parallactic slit angle, and therefore, the derived velocities might be potentially affected by systemic effects such as atmospheric dispersion. In order to retrieve a fit with $\chi^2_{\rm{red}}=1$, the required systemic uncertainty to be added in quadrature to the original uncertainties of the LRIS velocities is $11\, \rm{km \, s^{-1}}$. This produces $K_2= 218.9 \pm 2.2 \, \rm{km \, s^{-1}}$ and $\gamma = -21.3\pm 1.6 \, \rm{km \, s^{-1}}$ (Fig. \ref{figK2rv}, right panel), perfectly consistent with our previous fit. We also check the effect of leaving $T_{\rm{asc}}$ as a free parameter of the fit, and found similarly consistent results, as well as retrieved a $T_{\rm{asc}}$ matching that obtained from radio observations. We note that using different WD templates for the cross-correlation (with $T_{\rm eff}$ and $\log g$ values reasonably close to the selected one) also produce similar results.

We finally look for systematic effects in our results by analysing independently the datasets obtained with each instrument. On one hand, the higher resolution ESI data reveals $K_2= 221.1 \pm 2.5 \, \rm{km \, s^{-1}}$ and $\gamma = -21.3\pm 1.7 \, \rm{km \, s^{-1}}$ ($\chi^2_{\rm{red}}=0.96$ for 14 d.o.f.). On the other hand, the lower resolution LRIS data yields $K_2= 211.1 \pm 3.9 \, \rm{km \, s^{-1}}$ and $\gamma = -17.5\pm 3.2 \, \rm{km \, s^{-1}}$ ($\chi^2_{\rm{red}}=1.50$ for 21 d.o.f.). Both results are consistent within  $\sim 1\sigma $ and $\sim 1.5\sigma $, respectively, with those of the combined fit. 

As a result of the good agreement of all the previous fits, we employ hereafter the results obtained from the fit to the combined dataset with the LRIS uncertainties adjusted (Fig. \ref{figK2rv}, right panel). The combination of the derived $K_2$ with the orbital period $P_{\rm{orb}} $ and the pulsar projected semi-major axis ($x=0.5818172\pm 0.0000002)\, ls$), both derived from radio timing \citep{Lazaridis2009}, allow us to infer the mass ratio of the system:

$$q=\frac{M_{\rm{NS}}}{M_{\rm{WD}}}=\frac{K_2\, P_{\rm orb}}{2\pi\, x\,c}=10.44\pm 0.11$$ 

\subsection{Spectral classification of the WD companion}
\label{spectclas}

In order to perform the spectral classification of the WD, we average sets of normalised spectra obtained with the same configuration (same instrument, grism and slit width). Before combining these spectra, we correct each individual spectrum from its corresponding Doppler shift (as derived in Sec. \ref{RV}), setting them all in the WD companion reference frame. This yields averaged, null radial velocity spectra for LRIS and ESI instruments. 

\begin{figure}
\includegraphics[width=\columnwidth]{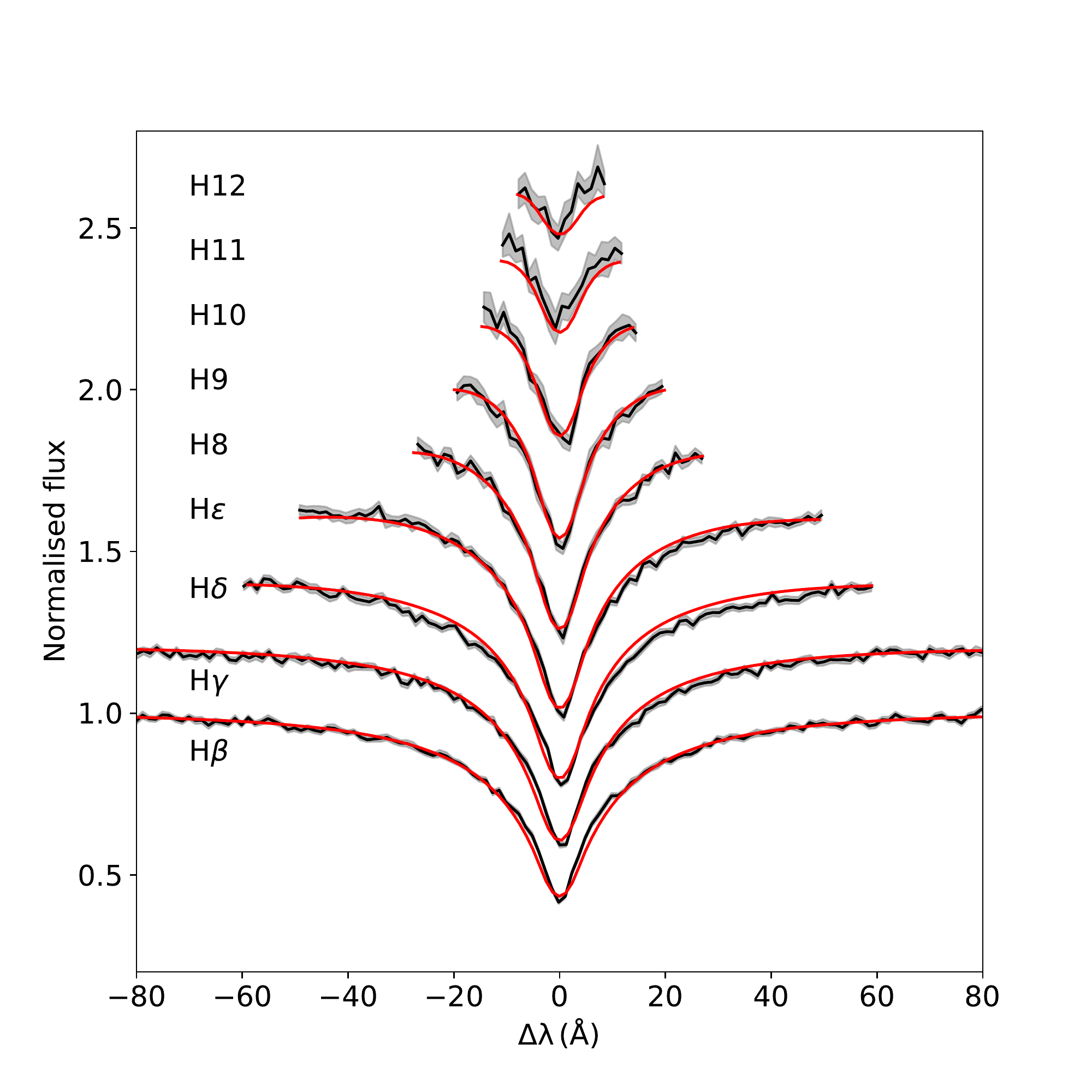}
\caption{WD Balmer series (from $H_{\beta}$ to $\rm H12$) in the LRIS averaged spectrum. A red, continuous line depicts the best-fit template after being broadened (through a Gaussian convolution) to the instrumental resolution of the observed spectra ($200\, \rm{ km\,s^{-1}}$). The grey shadow under the spectra defines the $1 \sigma$ uncertainty for the normalised flux.}
\label{figlinesLRIS}
\end{figure}

\begin{figure}
\includegraphics[width=\columnwidth]{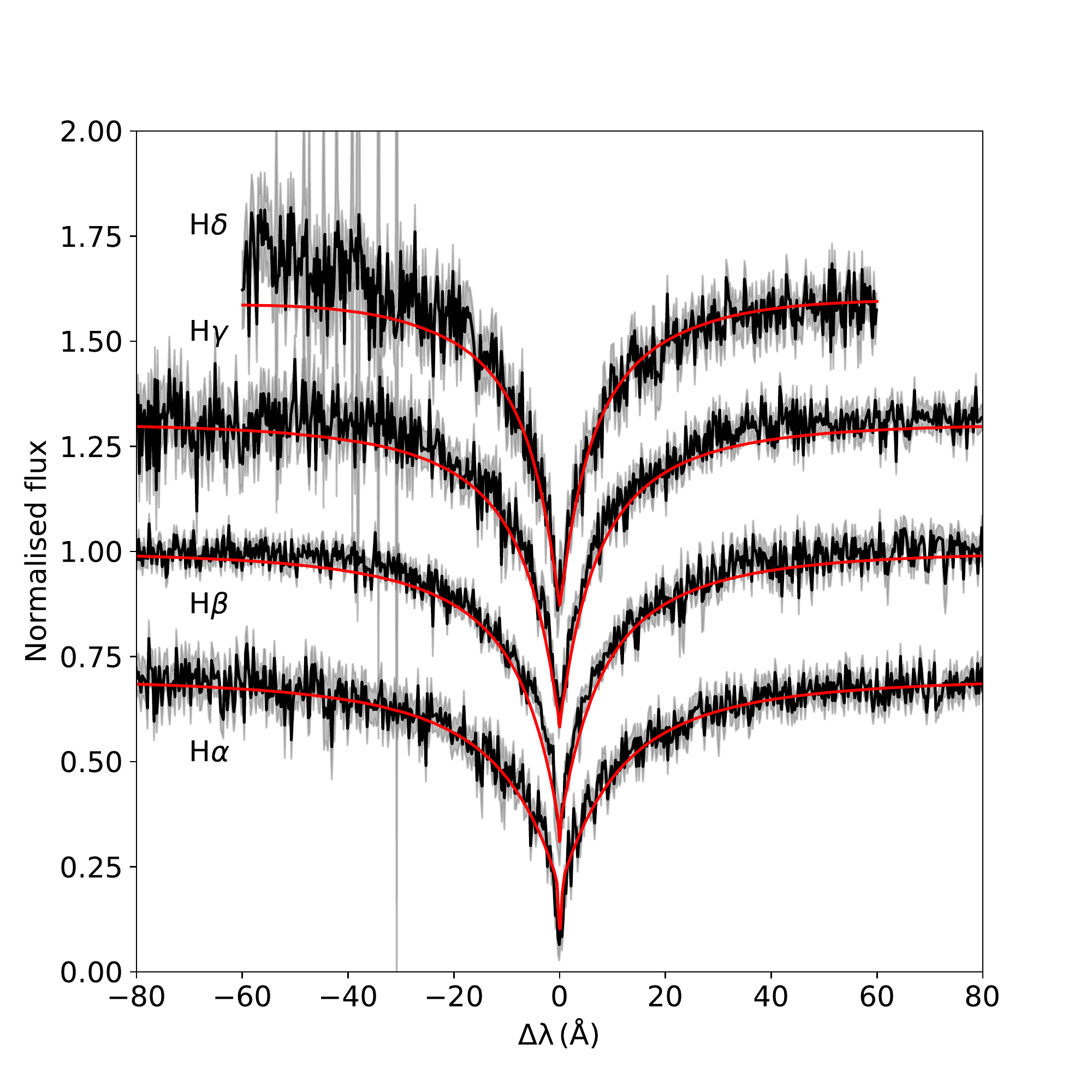}
\caption{Same as Fig. \ref{figlinesLRIS} but comparing with the ESI averaged spectrum. A red, continuous line depicts the best-fit template after being broadened (through a Gaussian convolution) to the instrumental resolution of the observed spectra ($45\, \rm{ km\,s^{-1}}$). The grey shadow under the spectra defines the $1 \sigma$ uncertainty for the normalised flux.}
\label{figlinesESI}
\end{figure}

We compare each averaged, normalised spectrum with the same WD templates that we employed for the radial velocity determination, covering a range of effective temperatures of $T_{\rm eff}=7250-8750 {\rm K}$ and surface gravities of $\log g = 5.75-7.75 $. The templates were also broadened to match the spectral resolution introduced by the instrument, via a convolution with a Gaussian function of full-width-at-half-maximum (FWHM) equal to that measured in the sky lines for each set ($200\, {\rm km\, s^{-1}}$ for LRIS and $45\,\rm { km\, s^{-1}}$ for ESI, see Tab. \ref{tab:Obslog}). To compare the observed data with each template, we apply a mask to only consider the WD features, that is, the Balmer lines. The LRIS spectrum covers transitions from $H_{\beta}$ to $\rm H12$ (see Fig. \ref{figlinesLRIS}), while the ESI spectrum only covers $H_{\alpha}$ to $H_{\delta}$ with sufficient signal-to-noise (SNR, see Fig. \ref{figlinesESI}). We analyse the averaged spectrum obtained with each instrument independently, but applying the same technique. 

To identify the template that best fits the data, we initially compute the reduced $\chi^2$ ($\chi_{\rm red}^2$= $\chi^2 /$ d.o.f.) resulting from the comparison between the averaged spectrum and each of the templates. This allows us to create a $\chi_{\rm red}^2$ map in the $T_{\rm eff}$ - $\log g$ parameter space, whose minimum corresponds to the best fitting template. In order to obtain the uncertainty of the best fit parameters, we normalised the map by its minimum value (which is equivalent to consider that the flux uncertainties have been underestimated). We then define the $1\sigma$ uncertainty as that corresponding to the 68.27 percentile. 

The resulting best fit parameters for the LRIS spectrum are $T_{\rm eff}=8540\pm 60 \,  {\rm K}$ and $\log g = 6.65 \pm 0.15$. Using the ESI spectrum instead, we obtain $T_{\rm eff}=8260\pm 80 \,  {\rm K}$ and $\log g = 6.1 \pm 0.4$. We find that both results are consistent within $\sim 2 \sigma$.

To better understand the uncertainties, we decided to recalculate the spectral classification by using a \textsc{python} script based on the routine \textsc{emcee} \citep{EMCEE2013}, an implementation of a Markov Chain Monte Carlo (MCMC) sampler. While the goodness of the fit is still fundamentally defined by the resulting $\chi^2$ from the comparison of the observed and template spectra, an MCMC approach should produce more reliable error bars on the derived parameters, as well as will allow for more control on the fit through the definition of priors. The grid of synthetic spectra was linearly interpolated to generate templates corresponding to any combination of $T_{\rm eff}$ and $\log g$ within the inspected range ($T_{\rm eff}=6500-8750\, K$, $\log g =5.75-7.75$), while spectral broadening was implemented through a convolution with a Gaussian kernel. We assume uniform priors on all these parameters, constraining $T_{\rm eff}$ and $\log g$ to the limits of the grid, and the instrumental broadening to be reasonably similar to the expected value. The method also includes a scale of the measurement uncertainties, equivalent to a prior that $\chi^2_{\rm red}=1$.

The best fit parameters when employing the LRIS spectrum are $T_{\rm eff}=8475^{+26}_{-21} \, {\rm K}$ and $\log g = 6.58^{+0.04}_{-0.04}$, where the results are expressed in terms of the percentiles of the posterior distributions (16, 50 and 84 \%), which show a Gaussian-like profile. These results are consistent with the previously derived values, but benefit from a higher precision, mostly because the interpolation in spectra is more accurate than the interpolation of $\chi^2$ values over the grid. On another vein, analysis of the ESI spectrum does not produce conclusive results, as the posterior distribution of the $\log g$ is found to be bimodal. This results in  $T_{\rm eff}=8154^{+50}_{-31} \,  {\rm K}$ and $\log g = 6.10^{+0.34}_{-0.11}$, values consistent with our previous analysis of this spectrum, but only marginally consistent with those from LRIS analysis.

On this topic, we note that the SNR of the ESI spectrum is lower than that of LRIS spectrum (due to the lower number of spectra combined), as well as that it covers a narrower range of features involving only the lower-order Balmer lines. Higher-order Balmer lines are more sensitive to the surface gravity parameter (see e.g., \citealt{Tremblay2009}), which might explain the better results obtained from the LRIS spectrum in spite of its lower resolution. Furthermore, normalization of the echellette spectra was more difficult, which may have affected the shapes of the very wide lines and thus the best fit. We propose that this prevented us from obtaining a reliable spectral classification of the higher resolution spectrum, and lead us to hereafter employ the parameters derived from the LRIS spectrum ($T_{\rm eff}=8475^{+26}_{-21} \,  {\rm K}$ and $\log g = 6.58^{+0.04}_{-0.04}$). 

It is important to also discuss possible systematic uncertainties in our inferred parameters, and in particular the determination of $\log g $. The so-called ``high $\log g $ problem'' has been noticed by several authors before (e.g., \citealt{Bergeron1990}) when analysing cold WDs ($T_{\rm eff}\lesssim 10000\, {\rm K}$). The spectroscopic models were found to overestimate $\log g $ when compared with the results from independent techniques (e.g., studying eclipsing systems, see \citealt{Gianninas2014} and \citealt{Kaplan2014}). On this topic, it is worth mentioning the work of \citet{Tremblay2015}. They performed a spectral classification of WDs using 3D atmosphere models, and compared their results with those obtained using 1D atmosphere models. They found coherent results between both sets of models except for the regime of $T_{\rm eff}= 8000-10000 \,  {\rm K}$, where the 1D models overestimate the value of $\log g$. They attribute this difference to an insufficient description of convection by the mixing length theory \citep{Bohm1958} in the 1D models, and they propose it as the origin of the ``high $\log g $ problem''. Using the correction functions presented in their work, our best estimates of the actual temperature and surface gravity of J1012 are $T_{\rm eff}=8362_{-23}^{+25} \,  {\rm K}$ and $\log g = 6.26_{-0.04}^{+0.04} $. We will use these values below.

\subsection{Rotational velocity of the WD}
\label{rotvel}

The rotational velocity for a companion star in a tidally locked circular orbit \citep{Wade1988}, rewritten in terms of known parameters for our particular case, is given by: 
$$v_{\rm rot} \sin i= \dfrac{2\pi\, R_2 \, \sin i}{P_{\rm orb}} $$

where $R_2$ is the WD radius and $i$ is the orbital inclination. As the orbital period is known from radio timing \citep{Lazaridis2009}, and assuming both the WD radius proposed in Sec. \ref{radiusWD} ($R_{\rm WD}=0.047_{-0.002}^{+0.003}\, R_{\rm \odot}$) and the inclination constrains from  Sec. \ref{NSmass} ($i=50\pm 2\, \rm deg$), the predicted rotational velocity (under assumption of the companion being tidally locked) is $v_{\rm rot} \sin i=  3.0 \pm 0.2  \, \rm{ km\,s^{-1}}$.

If the WD companion of J1012 is in a tidally locked orbit, the effect of the rotational velocity in the spectra (which translates into a broadening of the spectral lines) is probably well under our detection threshold even for our highest spectral resolution dataset ($45\, \rm{km\, s^{-1}}$, ESI spectra). Young WDs are expected to spin faster than co-rotation due to angular momentum conservation during the phase after the end of mass transfer, during which the remaining red-giant envelope contracts to the proto-white dwarf. On the other hand, older WDs might have slowed down due to tides. To cover all scenarios, we try to measure the spectral broadening for J1012 following an analogous technique to that employed for the spectral classification in Sec. \ref{spectclas}. For this particular purpose, the templates are broadened using the \textsc{python} routine \textit{rotbroad}, which allow us to emulate the line broadening for a particular rotational velocity value (as defined in \citealt{Gray1992}, and using a linear law to account for stellar limb-darkening). We inspected rotational velocities between $0-100\, {\rm km \, s ^{-1}}$ for our highest resolution ESI spectrum, and assumed typical limb-darkening coefficients for WDs (e.g., \citealt{Gianninas2013}). The minimum $\chi_{\rm red}^2$ is consistent with null rotational broadening, while the 68.27 percentile allow us to set an upper limit of $v_{\rm rot} \sin i < 60\, {\rm km \, s ^{-1}}$. Note that this limit is not as strict as one might hope in part because the line cores are slightly under-predicted even for zero rotational velocity.

\section{Discussion}
\label{discussion}

\subsection{The systemic velocity of J1012}
\label{dist3Dvel}

The spectroscopic analysis of Sec. \ref{RV} allowed us to derive a systemic radial velocity of $\gamma = -21.3\pm 1.6 \, \rm{km \, s^{-1}}$. \citet{Callanan1998} have previously reported a value of $\gamma = 44\pm 8 \, \rm{ km\,s^{-1}}$, based on an independent set of spectra (with spectral resolution $\sim 220  \, \rm{ km\, s^{-1}}$ and dispersion of $\sim 70 \, \rm{ km\, s^{-1}}$), which is inconsistent with our results. The value derived in our work have been obtained employing a spectroscopic database spread over two periods (separated $\sim 10$ years apart), and making use of two different instruments (of spectral resolution $\sim 200  \, \rm{ km\, s^{-1}}$ and $\sim 45  \, \rm{ km\, s^{-1}}$, respectively). The radial velocity analysis of the data obtained from both instruments showed consistent results both in the separate and joint fit (see Sec. \ref{RV}). For this reason, we favour the result obtained from our combined analysis, as the effect of systematic errors should be smaller.

\citet{Lazaridis2009} measured a transverse velocity of the system (referred to the solar system barycentre) from the proper motion of $v_{\alpha}=10.2\pm 1.0\, \rm{ km\,s^{-1}}$ and $v_{\delta}=101.5\pm 9.7\, \rm{ km\,s^{-1}}$. Combined with our derived systemic velocity, corrected for the small gravitational redshift of the white dwarf $v_{\rm GR}=2.2\pm 0.2\, \rm{ km\,s^{-1}}$ (i.e., $\gamma-v_{\rm GR}=-23.5\pm 1.6\, \rm{ km\,s^{-1}}$), the space velocity of the system is $v_{\rm space}=104.7\pm 9.5\, \rm{ km\,s^{-1}}$. This value is consistent with the results they obtained using the systemic velocity from \citet{Callanan1998}, as the main contribution to both the space velocity and the error budget comes from the transverse velocities. However, we note that the velocity vector direction is now different, which influences the derived evolutionary trajectory of J1012 calculated in \citet{Freire2011}. A recent work on J1012 (Ding et al. 2020, submitted to MNRAS; D20 hereafter) reported an updated distance value derived from radio parallax measurements performed with the Very Long Baseline Array (VLBA) of $d=828_{-18}^{+56}\, {\rm pc}$. Combined with the systemic velocity presented here, they show that the new predicted orbit for J1012 is perfectly consistent with those of the known MSP population.

\subsection{WD radius}
\label{radiusWD}

\begin{table}
\caption{Photometry compiled from the literature in the optical regime.}
\begin{threeparttable}

\bigskip

\centering
\begin{tabular}{l l l l l}
\hline
Band   & mag &   $f_{\nu} ({\rm mJy})$& Catalogue$^{a}$ \\
$\rm u'$   & $20.346\pm 0.056$  &  $ 0.0264\pm 0.0014$   & SDSS \\
$\rm g'$   & $19.610\pm 0.014$&   $  0.0520\pm  0.0007$  & SDSS \\
$\rm r'$   & $19.619\pm 0.017 $ &  $  0.0516\pm 0.0008 $   & SDSS \\
$\rm i'$   & $19.705\pm 0.026$&  $ 0.0476\pm 0.0011 $   & SDSS \\
$\rm z'$   & $19.830\pm 0.113$ &  $0.0425 \pm 0.0044$   & SDSS \\
g   & $19.605\pm 0.026$ &  $ 0.0523 \pm 0.0012$  & PS1 \\
r   & $19.653 \pm 0.013$ &  $ 0.0500 \pm 0.0006$  & PS1 \\
i   & $19.724\pm 0.013$ &   $0.0468  \pm  0.0005$ & PS1 \\
z   & $19.842 \pm  0.030$ &   $ 0.0420 \pm  0.0012$ & PS1 \\
y   & $19.79 \pm 0.05$ &   $0.0442 \pm 0.0020 $ & PS1 \\
$\rm G_{BP}$   & $19.67 \pm 0.05$  &   $ 0.0480 \pm 0.0022$  & Gaia DR2 $^{b}$  \\
$\rm G_{RP}$   & $19.39 \pm 0.06$ &   $ 0.0458 \pm 0.0027$  & Gaia DR2 $^{b}$  \\

\hline
 \\ 

\end{tabular}
\begin{tablenotes}
\item[a]{SDSS: Sloan Sky Digital Survey \citep{SDSS2015}; PS1: Pan-STARRS1 Survey \citep{PS12016}; Gaia DR2: Gaia Data Release 2 \citep{Gaia2018}.}
\item[b]{Gaia magnitudes are reported in VEGA system. The rest of magnitudes are reported in AB system.}
\end{tablenotes}
\end{threeparttable}
\label{tab:photo}

\end{table}

\begin{figure}
\includegraphics[width=\columnwidth,trim={2cm 0.5cm 3cm 2cm},clip]{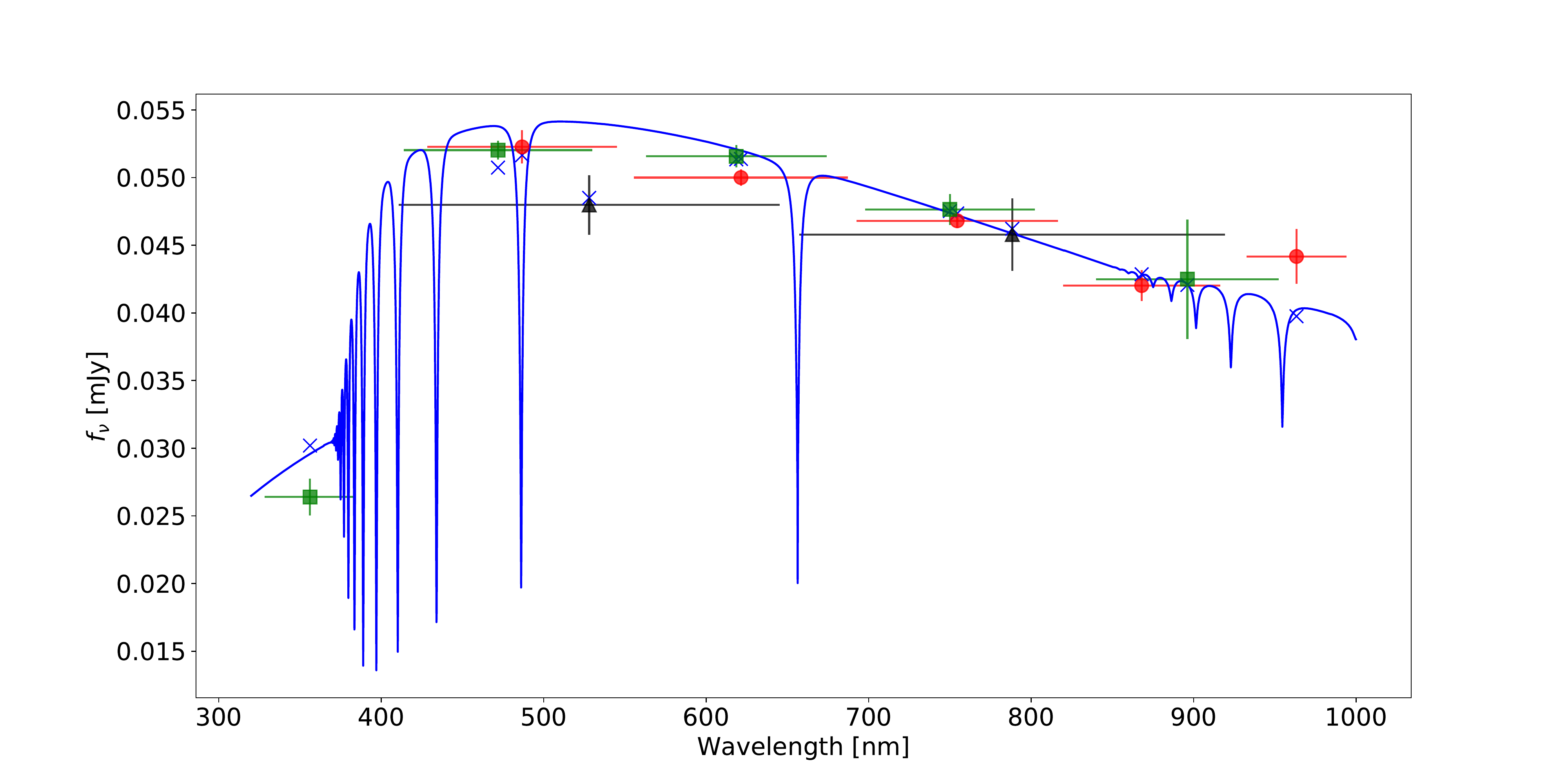}
\caption{Photometric SED in the optical range of J1012. Red circles, green squares and black triangles correspond to PS1, SDSS and Gaia DR2 photometry, respectively. A blue, solid line defines the best fit template flux-calibrated spectrum scaled to the distance measured for J1012. The blue extccrosses correspond to the theoretical SED calculated from the template in each of the bands.}
\label{figphoto}
\end{figure}

J1012 has been detected in a number of all-sky catalogues using a variety of photometric filters. The compilation of the available photometry in the optical range is shown in Tab. \ref{tab:photo}. We use this information to construct the observed spectral energy distribution (SED) for J1012 in the optical range (see Fig. \ref{figphoto}). On the other side, the synthetic template spectra employed in Sec. \ref{RV} determine the emitted Eddington flux (i.e. the flux density, $f_{\nu} =4 \pi f_{\rm{Edd}}$) in the optical range ($3500 - 9500$ \AA ) at the WD surface. By using the filter transmission curve associated with each observed photometric band (\citealt{PS1phot2012}, \citealt{SDSSDR72009} and \citealt{Gaiapassban2018} for PS1, SDSS and Gaia catalogues, respectively), we can construct the theoretical SED of the system at the WD surface from the synthetic spectrum using the formula:

$$f_{\nu}^{m}=\frac{\int_{0}^{\infty} f_\nu  S_\nu^{m} d\nu}{\int  S_\nu^{m} d\nu}$$

where $f_{\nu}$ is the synthetic spectrum flux density, $S_{\nu}^{m}$ is the transmission profile of the band $m$ (arbitrarily normalised, given the normalisation factor in this formula), and  $f_{\nu}^{m}$ is the total flux density integrated over the same photometric band $m$. 

As J1012 is well above the Galactic plane, interstellar extinction is negligible ($E(g-r)\lesssim 0.01$, \citealt{Green2018}). Therefore, the calculated theoretical SED at the surface of the WD and the observed SED from Earth must follow:
$$\dfrac{f_{\nu}^{ \rm{obs}}}{f_{\nu}^{\rm{surf}}}=\left( \dfrac{R_{\rm WD}}{d}\right) ^{2} $$
where $d$ is the distance to J1012 and $R_{\rm WD}$ the WD radius.

We find the best fit parameters ($T_{\rm eff}$ ,$\log g$, $R_{\rm WD}$, $d$) allowing us to match the observed and theoretical SEDs by using an MCMC sampler, following a method similar to that described at the end of Sec. \ref{spectclas} for the spectral classification. We assume uniform priors on the $T_{\rm eff}$  and the $\log g$ to constrain them to the available grid, as well as on $R_{\rm WD}$, which is constrained to be positive. We also impose a skewed Gaussian prior to the distance parameter given by the up-to-date most precise value $d=828_{-18}^{+56}\, {\rm pc}$ (D20).

The best fit values, expressed in terms of the percentiles of the posterior distributions (16, 50 and 84 \%) are: $T_{\rm eff}=8430_{-90}^{+80}\, K$, $\log g=5.91_{-0.11}^{+0.23}$, and $R_{\rm WD}=0.047_{-0.002}^{+0.003}\, R_{\rm \odot}$. The spectrum corresponding to the best fitting parameters is shown in Fig. \ref{figphoto}. The retrieved effective temperature is fully consistent with the value derived from the comparison of the absorption lines profiles (Sec. \ref{spectclas}). Given that each method relies on a different feature of the spectrum (either the continuum shape or the absorption lines profiles of the normalised spectrum), this reinforces our confidence on the derived value. The surface gravity, on the other hand, is pushed towards the lower end of the grid, and it is spread over a large range of values. This is a consequence the small influence that the $\log g$ parameter has in the continuum emitted light when compared with that of the $T_{\rm eff}$. The ratio between $R_{\rm WD}$ and the distance is well constrained in any case: it correlates fairly strongly with $T_{\rm eff}$, but hardly at all with $\log g$. The final uncertainty in $R_{\rm WD}$ is dominated by the uncertainty in the distance.

We now consider the conservative constraints on the NS mass of $1-3\, M_{\rm \odot}$ (see e.g., \citealt{Lattimer2012}), which combined with the derived mass ratio in Sec. \ref{RV} implies a valid range of the WD mass of $M_{\rm WD}=0.09-0.29\, M_{\rm \odot}$. Including this as a uniform prior to the $\log g$ in the previous analysis, the results are updated to: $T_{\rm eff}=8490_{-90}^{+90}\, K$, $\log g=6.13_{-0.09}^{+0.13}$, and $R_{\rm WD}=0.047_{-0.002}^{+0.003}\, R_{\rm \odot}$ ($\chi^2_{\rm{red}}=1.59$ for 9 d.o.f.).  The $\log g$ parameter is now better constrained, while both the effective temperature and the WD radius are perfectly consistent with the results previously determined. We also note that fixing the $\log g$ parameter to the corrected spectroscopic value ($6.26\pm 0.04$) in the photometric fit, we obtain consistent values of $R_{\rm WD}=0.046\pm 0.002 \, R_{\rm \odot}$ and $T_{\rm eff}=8520\pm 90\, K$. We conclude that these results provide additional support to the effective temperature derived in Sec. \ref{spectclas}, and allow us to employ hereafter the WD radius $R_{\rm WD}=0.047_{-0.002}^{+0.003}\, R_{\rm \odot}$.

\subsection{WD mass}
\label{massWD}

We have already established a range of viable masses for the WD companion ($0.09-0.29\, M_{\rm \odot}$) from the direct combination of NS mass conservative constraints and the measured mass ratio. Nevertheless, in order to retrieve the NS mass we need to obtain an independent measure of the WD mass. 

\subsubsection{Observational constraints}

We consider the corrected spectroscopic parameters of $T_{\rm eff}=8362_{-23}^{+25} \,  {\rm K}$ and $\log g = 6.26_{-0.04}^{+0.04} $ (Sec. \ref{spectclas}), which are consistent within $1.5 \sigma$ with the photometric results (Sec. \ref{radiusWD}). Employing the definition of surface gravity, as well as the measured WD radius ($R_{\rm WD}=0.047_{-0.002}^{+0.003}\, R_{\rm \odot}$), the resulting WD mass is $ M_{\rm WD}=0.15\pm 0.02\, M_{\sun}$.

\subsubsection{Evolutionary models}

A mass-radius relationship allows us to directly determine the mass of the WD from its measured radius. While such relationship is well defined for typical-mass WDs (e.g \citealt{Wood1995}), studying those in the ELM WDs regime usually requires extrapolation from the calculated models. Binary evolution must be considered to derive a more stringent constraint on $M_{\rm WD}$, as mass transfer is the underlying cause for both the low mass of the helium-core WD and the high NS spin frequency. The WD mass can be inferred by comparison of evolutionary models with the most precisely measured parameters for J1012: $T_{\rm eff}=8362_{-23}^{+25} \,  {\rm K}$, $P_{\rm{orb}}= 0.60467271355(3)\, \rm{d} $ and $R_{\rm WD}=0.047_{-0.002}^{+0.003}\, R_{\rm \odot}$.

\begin{figure}
\includegraphics[width=\columnwidth]{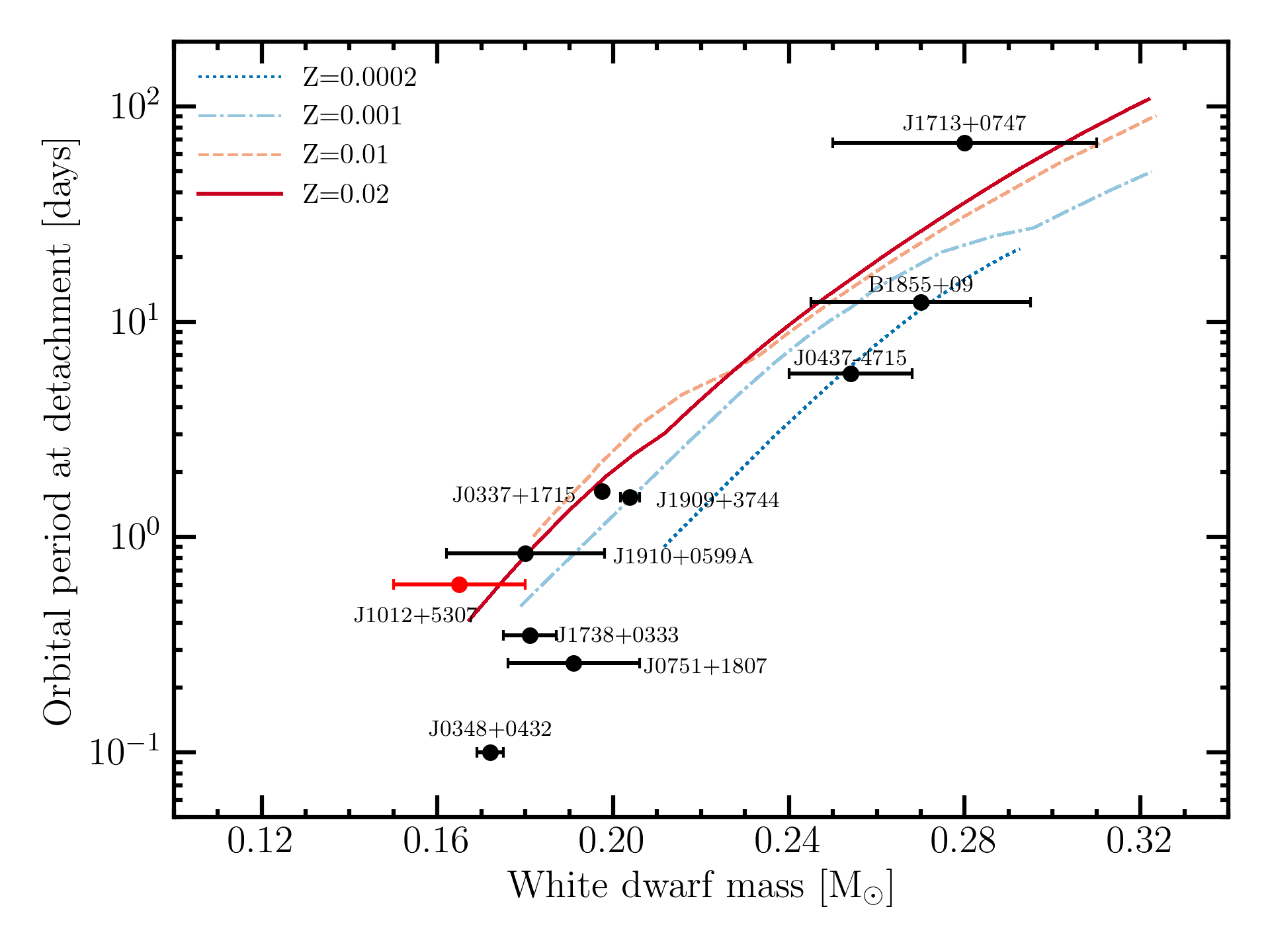}
\caption{Orbital period (at the end of the mass transfer phase) versus the mass of the proto-WD (this mass can be slightly different than the mass of the cooling WD as during the flashes the star can fill its Roche-lobe again several times). Overplotted with lines are the  numerical results obtained in \citet{Istrate2016}, for different metallicities. The circles represent various  He WD companions of MSPs (values taken from \citealt{Corongiu2012}). The red circle represents the constrains of PSR J012+5307 considered in this work.}
\label{fige.mass_period}
\end{figure}

It has been shown that there is a tight correlation between the degenerate-core mass of a red giant star and its radius \citep{Refsdal1971}. This relation depends primarily on the metallicity and to some extent on other stellar parameters, such as the mixing length $\alpha$ value adopted \citep{Han1998} or the initial stellar mass (e.g \citealt{Joss1987}). As a result, helium-core white dwarfs which are formed through stable mass transfer satisfy the so-called mass-period relation (e.g \citealt{ Tauris1999, Lin2011, Jia2014, Istrate2014, Istrate2016}). For white dwarfs smaller than $\lesssim 0.2\, M_{\rm \odot}$ (or orbital periods smaller than $\lesssim 2\, \rm{d}$), as a result of late case A  mass transfer, the mass-period relation shows some scatter. In this regime, the mass transfer is driven mostly by the loss of angular momentum, with magnetic braking being one of the most uncertain aspects of the evolution of these systems (e.g., \citealt{Van2019} and references therein). It was already pointed out that the classical formulation of the magnetic braking \citep{Rappaport1983} requires an extreme fine-tuning in the initial orbital period in order to explain the observed MSPs in very compact orbits ($2-9\, \rm{h}$, \citealt{Istrate2014}). We employ the evolutionary models presented in \citet{Istrate2016} for different metallicities to derive the mass-period relations shown in Fig.~\ref{fige.mass_period}.  They allow us to determine the mass of J1012 WD companion considering only its orbital period; for $Z = 0.02$, this would be 0.174~M$_{\odot}$, while a lower metallicity imply a more massive result. 

 On another vein, one can employ the same evolutionary models to fit instead the observed $T_{\rm eff}$ and $R_{\rm WD}$. The so-called \textit{basic} models required extrapolation to WD ages above that of the Universe in order to account for the observed parameters. Element diffusion in WDs (due to effects such as gravitational settling or chemical diffusion) affects significantly their cooling curve, and therefore the determination of their ages. Focusing on the family of models accounting for these effects, the derived age of the WD is better behaved ($\sim 10\, {\rm Gy}$), and produce a WD mass of $0.152\pm0.019\,M_{\rm \odot}$.

\begin{figure*}
\includegraphics[width=\columnwidth]{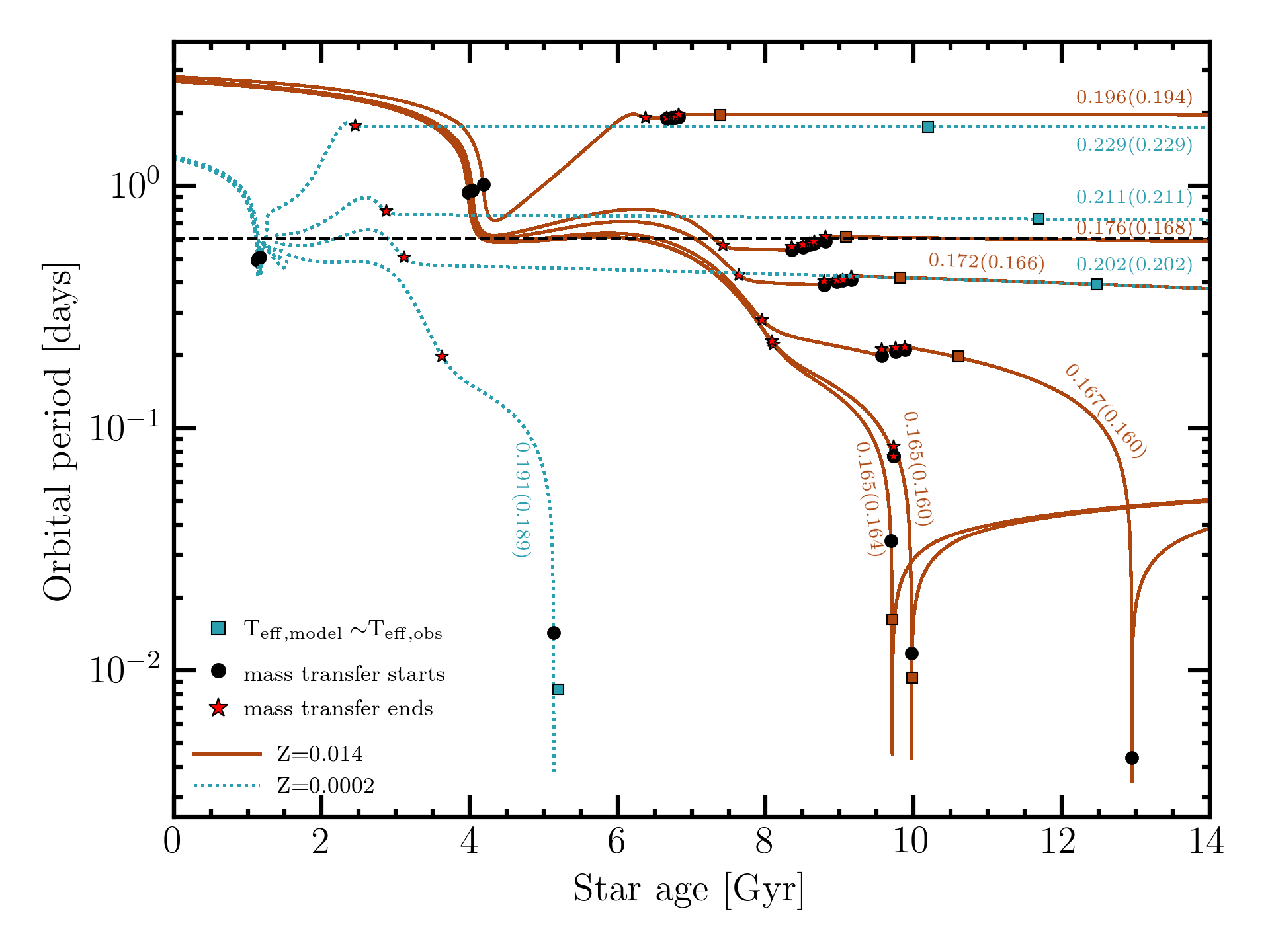}
\includegraphics[width=\columnwidth]{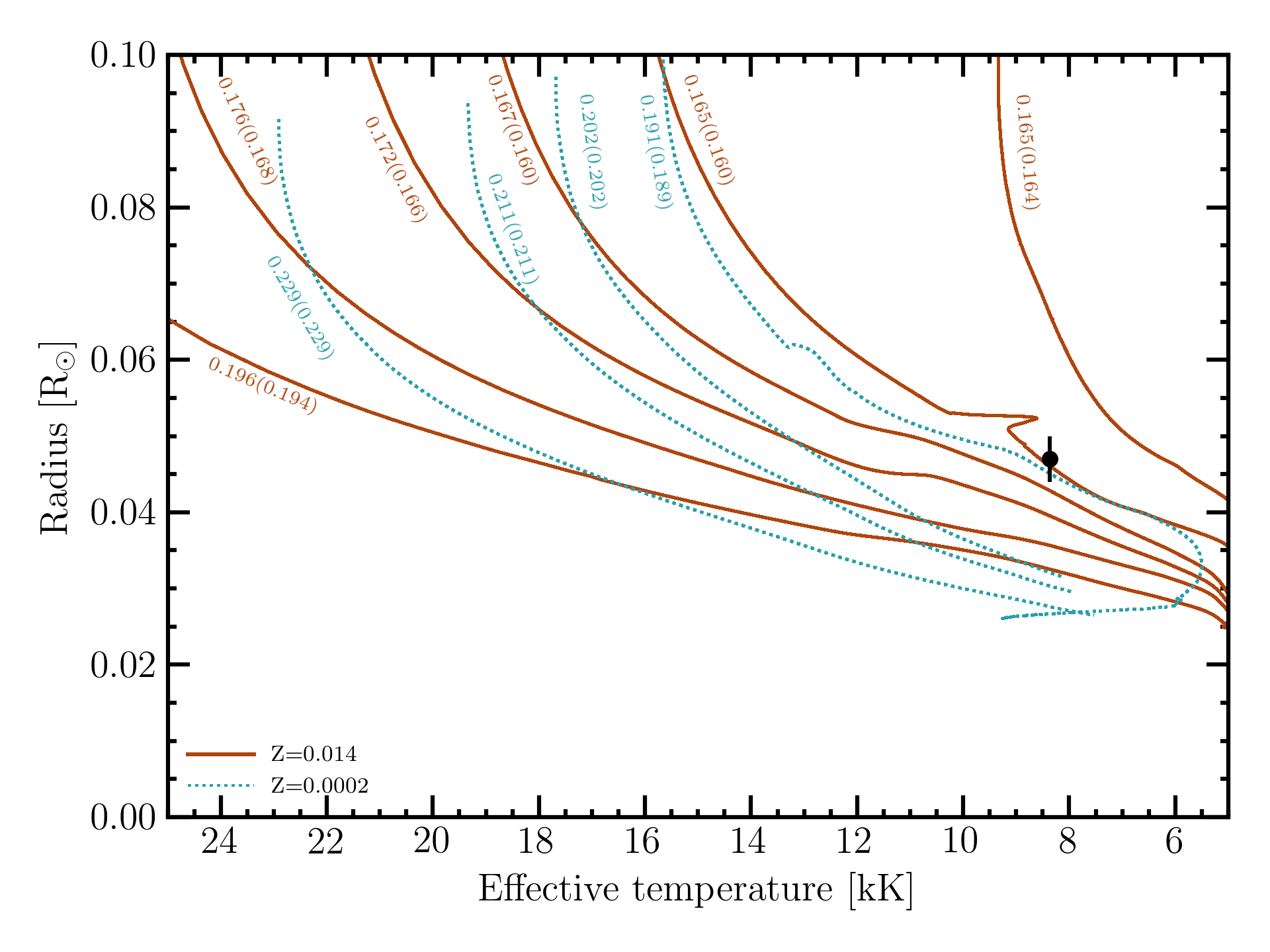}
\caption{Left panel: Orbital period versus age for evolutionary models with $Z=0.014$ and $M_{\mathrm{donor, initial}} = 1.1\, M_{\odot}$  (solid orange lines) and $Z=0.0002$  and  $M_{\mathrm{donor, initial}} = 1.4\, M_{\odot}$ (dotted turquoise lines). The black filled circles and the red filled stars determine the start and termination of the mass transfer phase. The  black dashed horizontal line marks the observed period for J1012. One should note that in the case when hydrogen flashes occur, several short mass  transfer episodes might follow. The squares mark the position in the diagram corresponding to the point when the WD reaches the observed effective temperature. Right panel: for the same set of models, the WD radius versus the effective temperature  during the cooling phase are shown. All models have as stop condition the requirement that the age of the system does not exceed 14 Gyr. The numbers next to the tracks indicate their mass at the end of the first mass transfer phase (same mass as shown in mass-period relation diagrams), while the value in parentheses represents the mass of the WD when the effective temperature matches the observed one.  The filled circle represents the values derived in this work.}
\label{figevol}
\end{figure*}

We note, however, that none of the inspected models were able to simultaneously account for the observed $P_{\rm{orb}} $ as well. In an attempt to achieve a solution that satisfies all the measured conditions, we computed additional evolutionary models using the binary stellar evolution code \textsc{mesa} \citep{Baxton2011,Baxton2013,Baxton2015} following the prescriptions defined in \citet{Istrate2014,Istrate2016}, using various metallicities, initial donor star mass as well as different mass transfer efficiency. While we found models matching the observed $T_{\rm eff}$ and $R_{\rm WD} $, none were able to reproduce simultaneously the observed $P_{\rm orb}$, always resulting in orbital periods well below the observed value. In Fig.~\ref{figevol} we show examples of the orbital evolution as well as the cooling evolution for two metallicities: Z=0.014 and a more extreme case of Z=0.0002. They allow us to to show the effect of this parameter in the evolutionary tracks, as well as to compare our results with previous works (which typically only consider solar metallicities). Considering only the cooling evolution of the WD, for Z=0.014 the mass required to explain the observed radius and effective temperature is $\sim 0.165 \, M_{\rm \odot}$. A similar result is obtained if we match instead the surface gravity ($\log g = 6.26 \pm 0.04$) and effective temperature determined spectroscopically ( $\sim 0.165 \, M_{\rm \odot}$). However, this mass regime  will result in a orbital period smaller than $\sim$ 0.3 days. The steep dependence of the mass with the orbital period is due to the underlying mass-period relationship, as shown in Fig. \ref{fige.mass_period}. To explain the observed orbital period, for Z=0.014 one would need a WD mass of $0.176 \, M_{\rm \odot}$.

At this point, it is important to note that all the masses discussed so far correspond to that of a WD just after the end of the initial Roche-Lobe overflow mass transfer. If no further mass-loss episodes from the WD were to occur, this would indeed correspond to the current WD mass. However, as shown by different authors, a significant envelope of hydrogen is still present in the WD after the main mass transfer event. Indeed, the absence of He transitions in the optical spectrum advocates in favour of this scenario for J1012. Parameters such as metallicity and the physics implemented in the evolutionary model (e.g., the magnetic braking) affect the thickness of the hydrogen layer, as well as determine if the hydrogen shell will experience stable or unstable burning phases. For the models with Z=0.014  considered in this work, a number of hydrogen flashes occur after termination of the main Roche Lobe overflow event (see Fig. \ref{figevol}). These short-lived epochs of hydrogen shell burning trigger mass losses from the WD, inevitably diminishing its final mass. Including this final ingredient, the mass of the WD would be $0.160 \, M_{\rm \odot}$ (if we consider $T_{\rm eff}$ and $R_{\rm WD}$); or else $0.168 \, M_{\rm \odot}$ (accounting only for the observed $P_{\rm orb}$ instead). Using $T_{\rm eff}$ and $\log g$ instead, the resulting mass would be $0.162 \, M_{\rm \odot}$. Again, no model was able to account simultaneously for all parameters (see Tab. \ref{tab:mass}).

\begin{table*}
\centering
\caption{J1012 WD mass derived from different methods and/or combinations of parameters.}
\begin{threeparttable}
\bigskip
\centering 
\begin{tabular}{l l l l}
\hline
$M_{\rm WD}\, (M_{\odot})$ & Method & Z & Reference  \\
\hline
$0.09-0.29$ & $1$  & & This work \\ 
$0.15\pm 0.02$ & $2$ & & This work \\
$0.174-0.200$& $3 {\rm a}$ & $0.02-0.0002$ & This work \\
$0.176\pm 0.001 \, (0.168\pm 0.001)$ & $3 {\rm a}$ & $0.014$ & This work \\
$0.165-0.192$ & $3{\rm b}$ & $0.02-0.0002$ & This work \\ 
$0.165\pm 0.001\, (0.160\pm 0.002)$ & $3{\rm b}$ & $0.014$ & This work \\ 
$0.159-0.180$ & $3{\rm c}$ & $0.02-0.0002$ & This work \\ 
$0.165\pm 0.001\, (0.162\pm 0.002)$ & $3{\rm c}$ & $0.014$ & This work \\ 
$0.16\pm 0.02$ & $4$ & & \citet{Callanan1998} \\ 
$0.16\pm 0.02$ & $4$ & & \citet{vanKerkwijk1996} \\ 
$0.17$ & $3{\rm c}$ & $0.00$ & \citet{Serenelli2001} \\ 
$0.165 - 0.170$ & $3{\rm b}$ & $0.01$ & \citet{Althaus2013} \\ 
$0.17\pm 0.01$ & $3{\rm c}$ & $0.01$ & \citet{Althaus2013} \\ 

\hline
\end{tabular}
\begin{tablenotes}
\item[] 1. Mass ratio ($q$) combined with conservative NS mass limits.
\item[] 2. Combination of $\log g$ with $R_{\rm WD}$.
\item[] 3.  Binary evolutionary models: ${\rm a/}$from $T_{\rm eff}$ and $P_{\rm orb}$ (i.e. mass-period relationship);  ${\rm b/}$from  $T_{\rm eff}$ and $R_{\rm WD}$, ${\rm c/}$from $\log g$ and $T_{\rm eff}$. Those WD masses obtained including the effect of hydrogen flashes are shown within parenthesis. The reported uncertainties on these mass values are conservative estimates based on model interpolation.
\item[] 4. Mass-radius relation, using  $\log g$.
\end{tablenotes}
\label{tab:mass}
\end{threeparttable}

\end{table*}

In order to reconcile these results, we consider below two scenarios:

\begin{itemize}

\item[\textbf{a/}] Some of the measured parameters that we employ to select the models are incorrect: We have already discussed about our confidence on the derived $T_{\rm eff}$, which seems quite robust as it has been independently derived through both the spectroscopic and photometric methods, and it is also consistent with previous works \citep{Callanan1998, vanKerkwijk1996}. The orbital period measurement proceeds from radio timing, a widely tested technique that is able to provide extremely precise ephemerides. The final parameter to consider would be the radius of the WD, which proceeds from the photometric fit (Sec. \ref{radiusWD}). This technique depends on the synthetic models for the WD spectrum, the observed SED and the measured distance to the system. The observed photometry seems quite robust (as it proceeds from independent catalogs, using a handful of different bands). The synthetic WD models have been widely employed/tested within many works, though maybe not as much for WDs in the ELM-regime due to the smaller known sample. While it seems they overestimate the $\log g$ of the WD in the particular regime J1012 lies in (as previously discussed), the effective temperature (which dominates the effect on the SED shape) seems much more reliable. The last remaining piece of the puzzle to assess is the distance to the system. If we were to trust the evolutionary models fitting both $P_{\rm orb}$ and $T_{\rm eff}$, that would require a WD radius ($R_{\rm WD}=0.038\pm 0.001\, R_{\odot}$) that, when combined with the optical photometry, yields a distance to the system of $710\pm 21 {\rm pc}$. The initially reported values from radio \citep{Lazaridis2009} and optical parallaxes \citep{Callanan1998} for the distance to J1012 were consistent within $<2\sigma$ with this value. However, the latest and more precise results from D20 are only marginally consistent ($\sim 3 \sigma$), making this scenario less plausible.

\item[\textbf{b/}] The evolutionary models are not complete: The mass-period relations suggest a smaller radius (i.e. higher mass) than the value derived from the observations. Assuming that the measured parameters are correct, this would point to an issue within the binary evolutionary models, or else, within the WD cooling models themselves (i.e. the mass-radius relation). \citet{Serenelli2001} modelled the evolution of helium-core WDs, and retrieved for J1012-like $T_{\rm eff}$ and $\log g$ a WD mass of $0.17\, M_{\odot}$. However, they did not take into account binary evolution, but instead obtained the white dwarfs by artificially removing mass from a 1.0 M$_{\odot}$. In a later work \citep{Althaus2013}, they modelled the full binary evolution, for an initial mass of 1.0 M$_{\odot}$ and Z=0.01. Their models provide a proto-WD mass (i.e., before hydrogen flashes occur) based on conservative values of $T_{\rm eff}$ ($8670 \pm 300\, K$) and $\log g$ ($6.34\pm 0.20$) of $0.17\pm 0.01\, M_{\odot}$. Using instead our more stringent parameters ($T_{\rm eff}$ and $R_{\rm WD}$), the favoured range of masses from their models is $0.165 - 0.170\, M_{\odot}$, which according to their tabulated values corresponds to a final orbital period of $0.35-0.43\, {\rm d}$, below the measured
value, and quite close to what is found from the models presented here.

A self-consistent evolutionary solution being able to explain the orbital evolution as well as the observed radius and effective temperature eludes us at this moment. As previously mentioned, in this regime there are several sources of uncertainties such as the initial donor mass, metallicity, mixing length parameter $\alpha$, stellar winds, efficiency of mass transfer, the prescription of magnetic breaking, just to name a few. 
For example, we note that using element diffusion during the evolution of the donor prior to the formation of the pre-ELM also influences the orbital period at detachment. All these input parameters not only affect the orbital evolution but also the mass of the envelope at the beginning of the pre-ELM phase, and more importantly, the total amount of hydrogen available. This in turn influences the number of flashes as well as the threshold for flash occurrence reflecting into a different amount of hydrogen at the beginning of the cooling track and therefore a slightly different cooling evolution. A detailed analysis of all these  uncertainties using J1012 as a benchmark will be addressed in a future work. 

A comparison of J1012 with similar binaries harbouring ELM WDs shows that the uncertainty on the derived parameters from evolutionary models is not exclusive of this system. The MSP+WD system PSR J1911-5958 shows a slightly higher effective temperature ($T_{\rm eff}=10090\, {\rm K}$) and a similar surface gravity ($\log g =6.44 \pm 0.05$), as well as a slightly larger orbital period ($P_{\rm orb}=20.6\, {\rm h}$) than J1012. The derived WD radius from the spectroscopic $\log g$ and mass-radius relations is not fully consistent with the results obtained considering the distance, $T_{\rm eff}$ and emitted flux \citep{Bassa2006}. The WD mass obtained from both methods ($M_{\rm WD}=0.18\pm 0.02\, M_{\odot}$ and $M_{\rm WD}=0.175\pm 0.010\, M_{\odot}$) is consistent thanks to the uncertainties involved. More recently, even the prototypical EL CVn, an eclipsing binary with a pre-He WD and a A-F type companion, has shown inconsistencies in the mass predicted by the evolutionary models with that calculated from an independent, dynamical analysis \citep{Wang2020}\footnote{However, for WASP+0247-25B \citep{Maxted2013}, another EL CVn-type system, an evolutionary solution that could explain simultaneously the measured orbital period, effective temperature and surface gravity as well as the observed pulsation periods was found \citep{istrate2017}.}. We can also point out the case of PSR J0348+0432, a MSP+WD system where the detection of the Shapiro delay has enabled the precise determination of many parameters. It possesses a higher effective temperature ($T_{\rm eff}=10120\, {\rm K}$) and similar mass ($M_{\rm WD}=0.172\pm 0.003\, M_{\odot}$) to J1012, but its orbital period ($P_{\rm orb}=2.4\, {\rm h}$) is significantly shorter \citep{Antoniadis2013} and the measured WD radius is significantly larger ($R_{\rm WD}=0.065\, R_{\odot}$).

It is also worth noting that ELM WDs lie in a region of the $P_{\rm orb}-M_{\rm comp}$ parameter space shared by another type of pulsar binaries known as `redbacks' \citep{Roberts2013}. The primary difference between these systems is that redbacks contain a semi-degenerate low-mass star having an optical spectrum reminiscent of that of a main sequence star. The reason why these two classes of compact binary pulsars overlap in properties and yet harbour different types of companions is still unknown. However, a study of the likely evolution of the redback PSR J2129$-$0429 revealed that it lies right at the boundary of evolutionary tracks separating systems evolving towards the position of J1012, which match the extension of the mass-period relationship, and other redbacks in which the companion has a lower-mass \citep{Bellm2016}.

\end{itemize}

Hereafter, we will employ as a conservative result for J1012 WD mass of $ M_{\rm WD}=0.165\pm 0.015\, M_{\odot}$, as it includes the different discussed scenarios, and it is fully consistent with the reported values from other works either employing different evolutionary models (e.g., \citealt{Althaus2013}) or using a extrapolated mass-radius relationship from more massive WDs (e.g., \citealt{vanKerkwijk1996,Callanan1998}).

\subsection{NS mass and orbital inclination}
\label{NSmass}

The analysis presented in Sec. \ref{RV} yields a radial velocity semi-amplitude for the WD companion of J1012 of  $K_2=218.9\pm 2.2\, \rm{km\, s^{-1}}$. \citet{vanKerkwijk1996} reported $K_2=280\pm 15\, \rm{km\, s^{-1}}$ based on a sub-sample of the low-resolution spectra presented in this paper (plus a single spectrum obtained at null orbital phase with a much worse spectral resolution, $\sim 740 \, \rm{km\, s^{-1}}$ and centred at the $\rm H_\beta$ wavelength). In a later work, \citet{vanKerkwijk2005} noted that their results were affected by a reduction error, and reported an updated value of $199\pm 10\, \rm{km\, s^{-1}}$, easier to reconcile with those shown here. An independent set of spectroscopic observations presented in \citet{Callanan1998} revealed $K_2=218\pm 10\, \rm{km\, s^{-1}}$, fully consistent with our result.

The combination of the newly derived $K_2$ with the pulsar radial velocity semi-amplitude ($K_1$) allowed us to derive the mass ratio of the system with unprecedented precision ($q=10.44\pm 0.11$), as well as produce mass functions for both components:

$$f_{M_{\rm{NS}}}=\frac{M_{\rm{NS}}\, \sin{i}^3}{(1+q^{-1})^2}=\frac{P_{\rm{orb}}\, K_2^3}{2\pi\,G}=0.66\pm 0.02  \, M_{\rm \odot}$$

$$f_{M_{\rm{WD}}}=\frac{P_{\rm{orb}}\, K_1^3}{2\pi\,G}= (5.782120 \pm 0.000006 ) \, \cdot 10^{-4} \, M_{\rm \odot}$$

\begin{figure*}
\includegraphics[width=2\columnwidth]{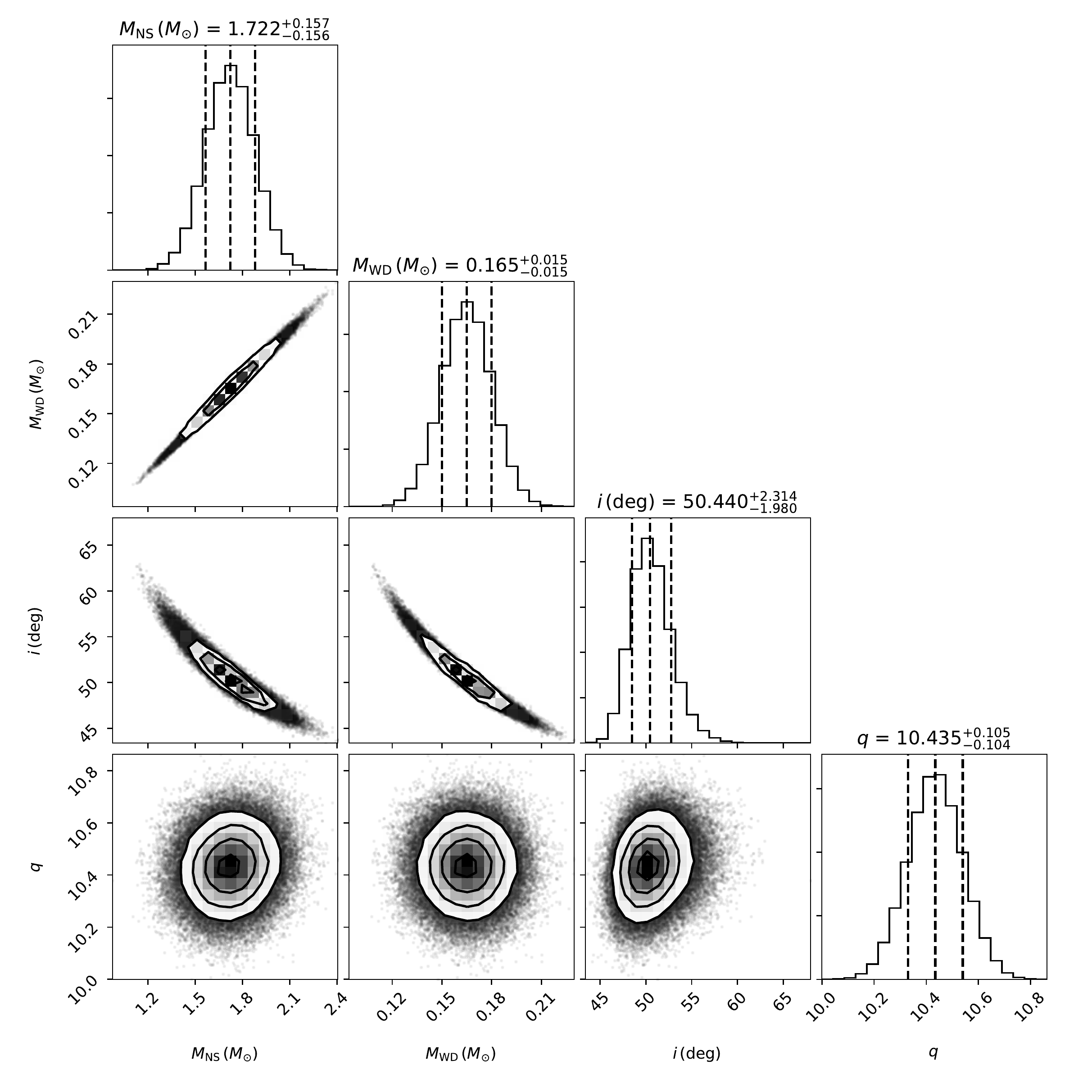}
\caption{Corner plot showing different dynamical parameters for J1012. They have been derived assuming Gaussian distributions for $K_1$, $K_2$ and $M_{\rm WD}$, using the values described along the text. The $1\sigma$, $2\sigma$ and $3\sigma$ contours are shown, and the percentiles 16,50 and 84$\%$ are plotted in the histograms.}
\label{figmass}
\end{figure*}

Combined with the known $q$, this imposes a lower limit to the NS mass which is not particularly constraining ($M_{\rm{NS}} >  0.79\pm 0.02\, M_{\odot}$) due to the unknown value of the orbital inclination ($i$). From the spectral classification of the WD, the photometric SED and the evolutionary models, we conclude in Sec. \ref{massWD} that the WD mass can be safely constrained to be $M_{\rm WD}=0.165 \pm 0.015 \, M_{\rm \odot}$. The addition of this final piece of information allow us to extract the remaining parameters of the system (see Fig. \ref{figmass}). While the inclination of the system results in $i=50\pm 2\, \rm deg$, the NS mass is constrained to $M_{\rm NS}=1.72\pm 0.16 \, M_{\rm \odot}$, revealing a compact object likely larger than the canonical value for NSs ($\sim 1.4 \, M_{\rm \odot}$, e.g., \citealt{Kiziltan2013}). Indeed, this result is perfectly consistent with the findings presented in \citet{Smedley2014}, where they compared the currently known population of MSPs with helium-core WD companions and, by assuming a random distribution of orbital inclinations, they concluded that they harbour NSs with masses of typically $\sim 1.6 \, M_{\rm \odot}$. Employing instead the WD mass constrains derived from the evolutionary models, the resulting NS mass would be either $\sim 1.66\pm 0.02 \, M_{\rm \odot}$ (using the measured $ R_{ \rm WD}$ and $T_{ \rm eff}$) or $\sim 1.76\pm 0.02 \, M_{\rm \odot}$ (using $P_{ \rm  orb}$ instead).

\subsection{Shapiro delay}

As noted in \citet{Lange2001}, the range $r$ of the Shapiro delay and the shape parameter $s$ are defined in terms of the dynamical parameters of the system as: 

$$r[{\rm{\mu s}}]=4.9255\,(M_{\rm{WD}}/M_{\odot})$$

$$ s\equiv \sin i = \left[ \frac{f_{M_{\rm{WD}}} (q+1)^{2}}{M_{\rm{WD}}}\right]^{1/3}$$

Our updated parameters produce $r=0.81\pm 0.07 \,  \rm{\mu s}$ and $s=0.77^{+0.03}_{-0.02}$, respectively. Using the re-parametrization from \citealt{Freire2010}:

$$\zeta = \frac{s}{1+\sqrt{1-s^{2}}};\qquad h_3=r\zeta^{3};\qquad h_4=h_3\zeta$$

We obtain $\zeta = 0.47^{+0.03}_{-0.02}$, $h_3=0.085^{+0.006}_{-0.005} \, \rm{\mu s}$ and $h_4=0.040^{+0.005}_{-0.004} \, \rm{\mu s}$.
\citet{NANOGrav2018} reported a non-detection of the Shapiro delay in this system, being the best-fit values of the orthometric Shapiro-delay parameters $h_3=0.02 \pm 0.07  \, \rm{\mu s}$ and $h_4=0.05 \pm 0.10  \,  \rm{\mu s}$. These are fully consistent with our results, which instead reveal the required precision to measure them.

\section{Conclusions}
\label{conclusion}

We present new spectroscopic observations of the MSP and ELM WD binary PSR J1012+5307, exhibiting the characteristic broad Balmer lines from its WD companion. The cross-correlations of the individual spectra with the selected template spectrum yield a WD radial velocity semi-amplitude of $K_2= 218.9 \pm 2.2 \, \rm{km \, s^{-1}}$, as well as a systemic velocity of $\gamma = -21.3\pm 1.6 \, \rm{km \, s^{-1}}$. Combined with the radio ephemerides determined from the pulsar study, it results in a precise mass ratio of $q=10.44\pm 0.11$. The spectral classification reveals a WD with $T_{\rm eff}=8362_{-23}^{+25} \,  {\rm K}$ and $\log g = 6.26_{-0.04}^{+0.04} $, while comparison with photometric observations of all-sky surveys produce a WD radius of $  R_{\rm WD}=0.047_{-0.002}^{+0.003}\, R_{\odot}$. Inspection of evolutionary models for ELM WDs lead us to propose a conservative WD mass of $  M_{\rm WD}=0.165\pm 0.015\, M_{\odot}$. This allow us to constrain the inclination of the system to $i=50\pm 2\, \rm deg$ and reveal a NS mass of $M_{\rm NS}=1.72\pm 0.16 \, M_{\rm \odot}$, a value slightly above the canonical one but fully consistent with the known population of MSPs. This work highlights the current limitations in our understanding of binary evolution and ELM WD physics. The uncertainty on the WD mass determination remains the main contribution to the error budget in determining the NS mass. In this regard, the detection of the Shapiro delay (and in particular of its $h_3$ parameter), as well as an even more precise photometric and parallax measurements would allow us to place tighter constrains on the physical parameters of J1012. An improved characterisation of this system will enable to test future revisions of the evolutionary models for ELM WDs.

\section*{Acknowledgements}

D.M-S. and R.P.B. acknowledge support from the ERC under the European Union's Horizon 2020 research and innovation programme (grant agreement No. 715051; Spiders). A.G.I thanks Gijs Nelemans for very helpful discussions and  acknowledges  support  from  the  Netherlands Organisation for Scientific Research (NWO).  DLK was supported by the NANOGrav Physics Frontiers Center, which is supported by the National Science Foundation award 1430284.
The authors wish to recognize and acknowledge the very significant cultural role and reverence that the summit of Mauna Kea has always had within the indigenous Hawaiian community.  We are most fortunate to have the opportunity to conduct observations from this mountain. This research has made use of the Keck Observatory Archive (KOA), which is operated by the W. M. Keck Observatory and the NASA Exoplanet Science Institute (NExScI), under contract with the National Aeronautics and Space Administration. This work made use of PyAstronomy. This research has also made use of the VizieR catalogue access tool, CDS, Strasbourg, France (DOI : 10.26093/cds/vizier). The original description  of the VizieR service was published in A\&AS 143, 23. This research has made use of the SVO Filter Profile Service (http://svo2.cab.inta-csic.es/theory/fps/) supported from the Spanish MINECO through grant AYA2017-84089. We  thank  the  anonymous  referee  for providing  helpful  comments to the manuscript.




\bibliographystyle{mnras}
\bibliography{MiBiblio}



\bsp	
\label{lastpage}
\end{document}